\pdfoutput=1%
\documentclass[10pt,a4paper]{article}%
\usepackage{opex3}
\usepackage{amsmath,amsfonts}
\usepackage{cite}
\usepackage{etex}
\usepackage[T1]{fontenc}
\usepackage{graphicx}
\usepackage[utf8]{inputenc}
\usepackage{microtype}
\usepackage{textcomp}
\usepackage[textsize=small,textwidth=35mm]{todonotes}
\usepackage{xspace}
\usepackage{subfigure}
\usepackage{hyperref}
\usepackage{layouts}
\usepackage{soul}
\usepackage[todos]{CMImacros}

\newlength{\figwidth}
\begin{document}
\title{Visualizing aerosol-particle injection for diffractive-imaging experiments}%
\author{\small Salah Awel,$\!^{1,2}$ Richard A.\ Kirian,$\!^{1,3}$ Niko Eckerskorn,$\!^{4}$
   Max~Wiedorn,$\!^{1,5}$ Daniel~A.\ Horke,$\!^{1,2}$ Andrei~V. \ Rode,$\!^{4}$
   Jochen~K\"upper,$\!^{1,2,5*}$ and Henry~N.\ Chapman$^{1,2,5*}$}%
\address{%
   $^1$Center for Free-Electron Laser Science, DESY, Notkestrasse 85,22607 Hamburg, Germany \\
   $^2$The Hamburg Center for Ultrafast Imaging, University of Hamburg, Luruper Chaussee 149, 22761 Hamburg, Germany \\
   $^3$Department of Physics, Arizona State University, Tempe, AZ, USA\\
   $^4$Laser Physics Centre, Research School of Physics and Engineering, Australian National University, Canberra, ACT 2601, Australia\\
   $^5$Department of Physics, University of Hamburg, Luruper Chaussee 149, 22761 Hamburg, Germany}%
\email{jochen.kuepper@cfel.de}%
\email{henry.chapman@cfel.de}%
\begin{abstract}
   Delivering sub-micrometer particles to an intense x-ray focus is a crucial aspect of
   single-particle diffractive-imaging experiments at x-ray free-electron lasers. Enabling direct
   visualization of sub-micrometer aerosol particle streams without interfering with the operation
   of the particle injector can greatly improve the overall efficiency of single-particle imaging
   experiments by reducing the amount of time and sample consumed during measurements. We have
   developed in-situ non-destructive imaging diagnostics to aid real-time particle injector
   optimization and x-ray/particle-beam alignment, based on laser illumination schemes and fast
   imaging detectors. Our diagnostics are constructed to provide a non-invasive rapid feedback on
   injector performance during measurements, and have been demonstrated during diffraction measurements at
   the FLASH free-electron laser.
\end{abstract}
\bibliography{string,cmi}%
\bibliographystyle{osajnl}%

\section{Introduction}
\label{sec:introduction}
The emergence of x-ray free-electron lasers (XFELs) has inspired the development of new
particle-injection instruments capable of delivering nano- and micro-particles to the intense
$0.1$--$5$~\um focus of a few-femtosecond duration x-ray beam. Single-particle diffractive imaging
(SPI) is among the methods that rely on the development of such particle-beam injectors, as it
requires a series of isolated molecules, viruses, cells or microcrystals to be directed across the
x-ray beam. Three-dimensional diffraction intensity maps can be constructed by assembling numerous
two-dimensional diffraction patterns from particles exposed in different
orientations~\cite{Bogan:AST44:i, Seibert:Nature470:78}. In this way, three-dimensional images can
be formed from reproducible targets. If successful, SPI will allow for the determination of
high-resolution structures of radiation-sensitive targets~\cite{Neutze:Nature406:752}, without the
need to grow large well-ordered crystals, which is often the principal bottleneck to macromolecular
structure determination.

In SPI experiments, it is important to precisely deliver the target particles to the most intense
region of the focused x-ray beam in rapid succession, since each particle is completely destroyed
through photoionization-induced damage processes~\cite{Chapman:NatMater8:299}. Liquid jets formed by
gas-dynamic virtual nozzles (GDVN)~\cite{DePonte:JPD41:195505}, aerodynamic aerosol
focusing~\cite{Bogan:NanoLett8:310}, or gas-phase supersonic jet/molecular beam
injectors~\cite{Kuepper:PRL112:083002} are among the most common techniques used to deliver
particles. For SPI work, gas-phase injectors are preferred since a surrounding liquid reduces
contrast and increases background scatter, which makes data analysis difficult, if not impossible.
Aerodynamical-lens-stack aerosol particle injectors (ALS)~\cite{Liu:AST22:293} are presently the
most common injector used for SPI experiments, which can create a collimated aerosol beam when
particles suspended in a carrier gas pass through a series of concentric apertures. Alternative
injectors, \eg, convergent-orifice nozzles, are also under development for SPI
experiments~\cite{Kirian:SD2:041717}.

During SPI experiments, aerosol injectors must be monitored frequently in order to maintain optimal
hit fraction and delivery efficiency, \ie, the fraction of x-ray pulses that intercept a particle
and the fraction of particles that are intercepted by an x-ray pulse, respectively. Particle-beam
diagnostics are important because XFEL facilities are costly to operate, and many samples are also
costly to obtain in significant quantities. X-ray diffraction patterns themselves are the ultimate
diagnostic of injection efficiency, but this diagnosis is limited by the XFEL pulse repetition rate,
detector readout rate, data processing rate, and availability of the x-ray source. It is desirable
to have complimentary real-time diagnostics that assist the injection optimization process, both
offline as well as online, during diffraction measurements. As we show below, direct visualization
of particle beams through laser illumination is a simple yet powerful means to optimize injection
efficiency. In addition to improving SPI experiment efficiency, imaging diagnostics can greatly
accelerate the development of new aerosol injector schemes.

Aerosolized nanoparticles are not easily visible, and particle injection environments are not always
easily accessible for probing due to ancillary measurement tools. Therefore, \emph{in-situ}
diagnostics can be challenging to implement within existing x-ray diffraction apparatuses. Early
experimental work utilized greased plates onto which aerosol particles
adhere~\cite{Rao:JAerosolSci24:879, Benner:JAerosolSci39:917}, allowing the transverse particle beam
profile to be estimated. This method is commonly employed in SPI experiments, however such particle
depositions, examined under a microscope, are not easy to interpret quantitatively. In the context
of SPI work, the first detailed experimental characterization of aerodynamically focused particles
was carried out by Benner \etal~\cite{Benner:JAerosolSci39:917}. Here, particle velocities and
positions were determined from the image charges of particles transmitted through a metal tube.
Aerosol beams have also been directly imaged in the past~\cite{Fuerstenau:JAerosolSci25:165,
   Farquar:JAerosolSci39:10}, but so far the great utility of this approach that we emphasize here
has not been integrated into SPI experiments. More generally, the determination of particle-laden
flow fields has been studied extensively within the field of particle image velocimetry (PIV) and
its variants~\cite{Adrian:PIV:2011, Willert:ExpFluids10:181}.

In this paper, we present simple direct optical imaging diagnostics for online monitoring of
particle injection during XFEL experiments, as well as for general aerosol beam characterization and
injector optimization. We have utilized both continuous-wave (CW) and pulsed nanosecond illumination
along with high-speed cameras, and nearly real-time analysis software, that can measure particle
speeds, injector transmission efficiency, and projected particle beam density profiles. We have also
implemented an in-vacuum inverted microscope for imaging particles that adhere to a gel.

\section{Theory and Background}
\label{sec:TheoryBackground}
\subsection{Efficiencies of single-particle imaging}

Direct optical imaging can reveal most of the key parameters needed to optimize SPI sample
injection, which we discuss here. The first key parameter is known as the ``hit fraction'' or ``hit
rate'', and is equal to the fraction of x-ray pulses that intercept a particle. For femtosecond XFEL
illumination, this quantity depends on the instantaneous projection of the particle number density
along the x-ray beam path, and can be approximated as $H\approx{}fT\sigma/(vd)$, where $f$ is the
rate at which particles enter the injector, $T$ the injector transmission efficiency (ratio of the
number of particles that enter and exit the injector), $\sigma$ the effective illumination area that
produces useful diffraction, $v$ the velocity of the particles, and $d$ the particle beam diameter.
In this formulation, we assume that the particle beam has a diameter smaller than the depth of focus
of the x-ray beam, which is almost always satisfied in SPI experiments.
We also assume that the x-ray beam diameter is significantly smaller than the particle beam
diameter. We assume that $H < 1$, since x-ray diffraction patterns containing multiple particles
illuminated simultaneously tend to complicate the diffraction analysis.

Another key parameter is known as the ``delivery efficiency'', equal to the fraction of consumed
particles that are intercepted by an x-ray pulse. Delivery efficiency for continuous flow of
particles can be approximated as $\epsilon\approx{}HF/f$, where $F$ is the XFEL pulse repetition
rate, and it is assumed that $F\ll{}f$. Notably, a hit fraction of $H\approx1$ can be achieved while
having delivery efficiency $\epsilon \ll 1$.

In order to score a higher hit fraction and delivery efficiency, one typically needs to find an
optimal compromise between the three parameters $v$, $d$, and $T$. Importantly, for a given injector
geometry, it might not be possible to vary these parameters independently of each
other~\cite{Wang:AST40:320}. Different types of injectors can also introduce tradeoffs -- for
instance, a convergent-orifice injector can create a tightly focused particle beam that approaches
the size of micro-focused x-ray beams, but apparently produces particles with greater speeds than
typical ALS injectors~\cite{Kirian:SD2:041717}. Tightly focused beams \emph{necessitate} the use of
an \emph{in-situ} direct imaging system since one would otherwise need to perform a
three-dimensional scan of the x-ray beam in order to properly position the interaction region,
whereas a collimated particle beam requires only a two-dimensional scan.

In general, independent of the type of injector used, the aerosol beams we consider here are
composed of fast, nearly-unidirectional, and sparsely placed small particles confined to a narrow
beam in a low-pressure environment. Typically, on the order of $10^{7}$ particles enter the injector
per second and expand into the vacuum with a speed that can reach several hundred m/s. This leads to
hit fractions well below 0.1\% for current injectors and nano-focused x-ray beams, thus rendering
x-ray diffraction-based diagnostics inefficient, highlighting the need for complementary
rapid-feedback diagnostics.

\subsection{Direct side-view particle imaging schemes}
\label{sec:imaging-schemes}
We can classify the direct side-view imaging of particles presented here into three regimes,
principally identified by three characteristic times: $\tau=d/v$ -- the time it takes for a particle
with velocity $v$ to move over its diameter $d$, exposure time $t_\text{exp}$ --
the camera integration time or duration of the illumination pulse, and $t_\text{fov}$ -- the time
taken for a particle to move across the full field of view (FOV). $d$ is the diffraction-limited
spot size of the particle if the particle is smaller than the resolution limit of the of the imaging
system.

\subsubsection{``Long exposure'' imaging}
\label{sec:long-exposure-imaging}
In the ``long exposure'' mode, the particle beam is illuminated either with a continuous or pulsed
light source with a very long exposure time ($t_\text{exp}\gg{}t_\text{fov}$) on the
camera~\cite{Fuerstenau:JAerosolSci25:165}. This mode does not allow for the determination of
particle velocities, but is straightforward to implement with relatively inexpensive equipment. In
many cases the resulting integrated image intensity is directly proportional to the projection of
the particle density along the optical axis. However, since elastic scattering in both the Mie and
Rayleigh regime scales exponentially with particle diameter (for Rayleigh scattering, the intensity
scales with the sixth power of particle diameter), one must ensure that all particles are of the
species of interest, and not aggregated clusters of particles (for example) that would tend to
dominate the intensity profile of the image.

\subsubsection{``Streak'' imaging}
\label{sec:streak-imaging}
Visualizing \emph{individual}, fast-moving, sub-micrometer particles requires
$t_\text{exp}<t_\text{fov}$, such that the entire image is contained within the field of view. In
the ``streak'' imaging mode, $t_\text{exp}$ is chosen such that particles appear as streaks across
the imaging plane $(\tau<t_\text{exp}<t_\text{fov})$. If $t_\text{exp}$ is known and the entire
streak is contained in the image, the velocity can be determined from the streak length. If the
particle density is sufficiently low to avoid overlapped particle images, the number density of
particles can also be determined by analyzing the intensity centroid of each streak. Ideally, the
illumination source should have a well-defined top-hat temporal profile as well as uniform spatial
intensity profile. This can be achieved with CW lasers, provided a fast shutter is available for
either the laser or the imaging device. Longer streak lengths lead to better measurement accuracy,
but also increase the chance of particle streaks overlapping and of streaks that partly fall out of
the field of view. The optimum $t_\text{exp}$ should be chosen according to these two factors.

\subsubsection{``Snapshot'' imaging}
\label{sec:snapshot-imaging}
In the ``snapshot'' imaging mode, when $t_\text{exp}\ll\tau$, point-like particle images are
produced on the detector, mitigating motion blur~\cite{Thoroddsen:ARFMech40:257,
   Versluis:ExpFluids54:1}. For example, particles moving at $v=200$~m/s with $d=1~\um$ require a
5~ns exposure time to freeze the motion. The snapshot image can be achieved with short camera
integration times or short illumination sources, \eg, pulsed lasers, flash lamps, or spark
discharges~\cite{Adrian:PIV:2011}. We note that in the cases of streaked and snapshot imaging modes,
one can determine particle positions at a resolution better than the resolution of the optical
system through intensity centroid analysis, akin to super resolution microscopy molecule
localization techniques~\cite{Betzig:Science313:1642, Henriques:Biopolymers95:322}. The snapshot
imaging mode has several advantages over the long-exposure imaging mode: it enables straightforward
quantitative determination of particle beam density, and in principle one can infer particle volumes
through integrated scattering intensity, if the system is well calibrated. The velocity and
acceleration of particles can also be measured from snapshot images with the use of multiple
exposures with known delays, provided that all particle images appear in the same field of
view~\cite{VanDerBos:PhysRev1:014004, Adrian:PIV:2011}.

Provided that detector readout noise is not significant, snapshot imaging maximizes the
signal-to-noise ratio (SNR) since unnecessary exposure time is avoided, and all scattered light
entering the optics is focused to a single resolution element. Imaging based on continuous
illumination and short camera integration time usually suffers from lower signal levels compared
with pulsed illumination (assuming similar average optical power). This is due to the fact that in
the latter case the intensity of a particle image is fixed by the intensity of the illumination,
whereas in the former case only a small fraction of the CW laser power is used to illuminate the
particle, \ie, most of the laser power is unused~\cite{Thoroddsen:ARFMech40:257, Adrian:PIV:2011}.

\subsection{Transverse-plane particle imaging}
\label{sec:Invaccum}
Simple imaging of the transverse profile of the particle beam can be achieved with the help of a
flat, sticky surface placed transverse to the particle beam propagation. The particle beam diameter
can be roughly estimated from the deposition of particles on the plate, imaged either directly
\emph{in-situ}, or by analyzing the deposition under an external
microscope~\cite{Rao:JAerosolSci24:879, Benner:JAerosolSci39:917}. Unlike side- view imaging
techniques, this does not contain any information regarding particle dynamics, but nonetheless gives
useful and rapid feedback on the performance of an injector. For instance, asymmetry of the particle
beam in the transverse plane is difficult to observe with side-view imaging, but can be observed
easily using this technique.

\subsection{Laser scattering intensity}
Imaging sub-micrometer particles through elastic scattering raises considerable concerns regarding
scattering intensity at the detector. As we show below, a relatively modest setup can be used to
image particles with diameters of a few hundred nanometers, where Mie scattering dominates. Mie
scattering theory is typically applied for particle diameters down to approximately one tenth of the
scattering wavelength, below which the simple Rayleigh theory becomes applicable. The latter is
generally considered valid for the case $d\pi/\lambda<1$, where $d$ is the particle diameter and
$\lambda$ the wavelength of light. For a wavelength of 532 nm, this corresponds to $d\sim$170 nm. In
the regime $\sim 50-170$~nm both Mie and Rayleigh theories can be considered valid and yield
comparable scattering cross-sections (see below). However, they differ significantly in theoretical
treatment. Mie theory is based on an infinite series of spherical partial waves to describe
scattering, whereas the Rayleigh approximation can be summarized as a single analytical expression.
The former calculates the (complex) scattering phase functions, and therefore yields a directional
scattering dependence, while Rayleigh theory assumes an isotropic scattering distribution (apart
from a polarization correction). In the following basic theoretical treatment we focus on Rayleigh
scattering theory, due to its mathematical simplicity and because it is a valid approximation in the
size range of typical biological molecules.

\begin{figure}
   \centering
   \includegraphics[width=0.6\linewidth]{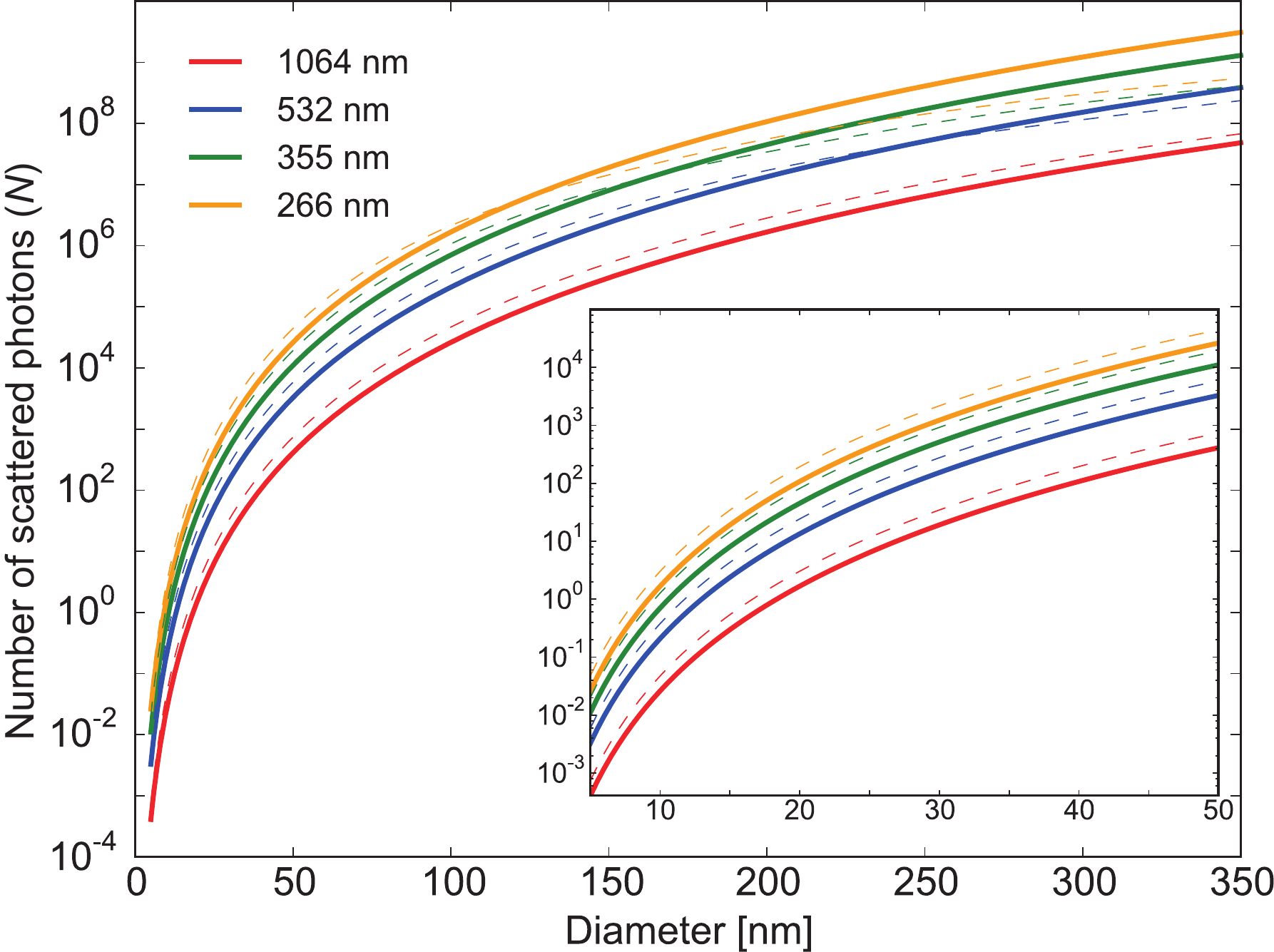}
   \caption{Total number of scattered photons as a function of particle diameter for several
      wavelengths. Solid lines are calculated using the Rayleigh formalism, dashed lines are
      calculated from Mie theory. The calculation is done for 100~mJ pulses ($N_0\approx10^{17}$
      photons) focused to a top-hat spatial intensity profile with diameter $\omega_0=1$~mm.}
   \label{fig:rayleigh}
\end{figure}
The total Rayleigh scattering cross section for a sphere of diameter $D$ and relative permittivity
$\epsilon$ is~\cite{Miles:MST12:R33}
\begin{equation}
   \sigma =\frac{8\pi^5d^6}{3\lambda^4} \left(\frac{\epsilon-1}{\epsilon+2}\right)^2 \;
\end{equation}
For a beam of diameter $g$ and pulse energy $E_0$, the scattered energy is $E = 4E_0\sigma/\pi g^2$,
and the number of scattered photons is $N=E\lambda/hc$, where $h$ is Planck's constant, and $c$ is
the speed of light. Thus, we have
\begin{equation}
   N = \frac{32E_0\pi^4d^6}{3\lambda^3g^2hc}\left(\frac{\epsilon-1}{\epsilon+2}\right)^2 \;
\end{equation}
The relative permittivity for proteins can vary significantly~\cite{Li:JCTC9:2126}, but
$\epsilon\approx2\text{--}4$ is a reasonable assumption; polystyrene has $\epsilon\approx2.6$.
\autoref{fig:rayleigh} shows total scattering calculations for the Rayleigh and Mie regime for
different particle diameters and laser wavelengths for the case $\epsilon=2.6$. We must reduce the
total scattered photon number $N$ according to the fraction of photons observed. This results in a
number of photons $N_\Omega$ captured in the solid angle $\Omega$ of the optical system. Neglecting
polarization factors, we obtain
\begin{equation}
   N_\Omega = \frac{16E_0\pi^4d^6}{3\lambda^3g^2hc} \left(\frac{\epsilon-1}{\epsilon+2}\right)^2 (1-\cos\theta)
\end{equation}
where $\theta$ (measured from the optical axis of the imaging system) is the maximum scattering
angle collected by the optical system (the numerical aperture is defined as $\text{NA}=sin\theta$).
This provides a lower bound for experiments with the polarization axis of a linearly-polarized laser
perpendicular to the optical axis of the imaging system.

The SNR of an imaging system depends on several factors. Since the typical size of single particle
scattered intensity spans very few pixels on the detector, a pixel will collect approximately
$N_\Omega$ photons from a particle. If the dominant noise sources of the imaging chip are the dark
current, readout noise, background photons, and Poisson noise, the signal-to-noise ratio can be
expressed as
\begin{equation}
   SNR = \frac{N_\Omega Q}{\sqrt{N_\Omega Q + N_b Q + N_d + \sigma_r^2}}
\end{equation}
where $Q$ is the quantum efficiency of the chip (number of electrons per photon), $N_d$ is the mean
number of dark current electrons, $\sigma_r$ is the RMS readout noise (in number of electrons), and
$N_b$ is the number of background photons per pixel. This estimate assumes that all photons
collected by the objective are directed to a single pixel. As an example, the camera utilized in our
measurements (Photron SA4) contains a CMOS chip that has a readout noise of 38 electrons, and a
quantum efficiency of about 33\% at 530 nm. Assuming that background photon levels can be reduced to
nearly zero, a minimum of about $38/0.33 \approx 120$ photons per pixel would be required to obtain
a SNR of 1 with this chip. Factoring in the collection angle of the optics, we can roughly estimate
that particles down to about 50~nm could likely be imaged with this detector. Smaller particles may
require the use of single-photon detectors, such as EMCCD and SPAD detectors~\cite{Fowler:SPI:159,
   Krishnaswami:OptNano3:1}.

\section{Experimental setup}
\begin{figure*}
   \centering
   \includegraphics[width=1\linewidth]{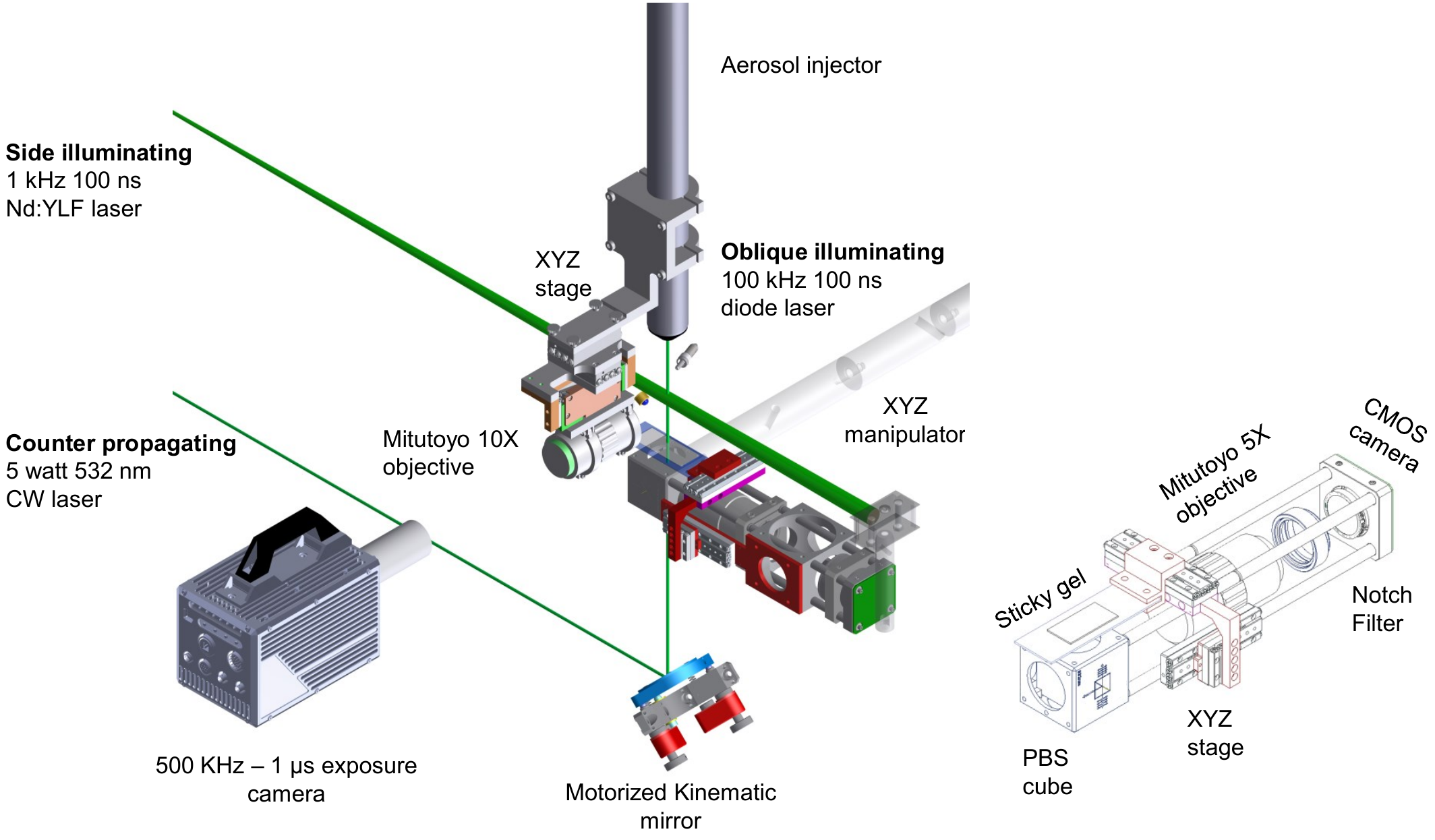}
   \caption{A schematic diagram of the basic direct aerosol imaging set-up. The in-vacuum microscope
      assembly is show to the right. }
   \label{fig:setup}
\end{figure*}
Our experiment is constructed within a vacuum chamber that hosts an aerosol injector and, in some
cases, x-ray diffraction detectors. For nebulization we use a gas-dynamic virtual nozzle
(GDVN)~\cite{DePonte:JPD41:195505, Seibert:Nature470:78}. The aerosol stream is delivered either by
an ALS injector (Uppsala University, Sweden) or by a convergent-nozzle
injector~\cite{Kirian:SD2:041717}.

Side-view imaging in all three modes is implemented using a high-speed imaging configuration based
on a high-frame-rate camera or on pulsed-illumination, as shown in \autoref{fig:setup}. Imaging in
the transverse plane is achieved with an inverted in-vacuum microscope that views particles as they
adhere to a glass microscope slide coated with a sticky purified gel film (TELTEC, P/N
DGL-20/17-X8).

\subsection{Side-view imaging configuration}
\label{sec:SideImaging}
The key components in our side-imaging system are a high-frame-rate camera, both pulsed and CW
illumination lasers, and imaging optics optimized for either a wide field of view or a high
magnification. We discuss these components and their configurations below. We generally work in a
quasi-dark-field imaging mode, where images are formed from scattered light without allowing the
direct beam to enter the optical system. For wide-field views, we use a long-working-distance (LWD)
microscope (Infinity model K2, working distance 225--300~mm, depth of focus (DOF)
$\approx\!100~\um$, magnification 2.13, and FOV $11.7\times11.7\text{~mm}^2$) mounted outside of the
vacuum chamber. For high-magnification views, a $10\times$ infinity-corrected objective (Mitutoyo,
working distance 38~mm, DOF 3.5~\um, magnification $28$, FOV $850\times850~\um^2$) is used, mounted
on a three-axis motorized stage inside the vacuum chamber. Switching between these two
configurations only involves swapping in/out the K2 objective and translating the high-magnification
objective into position. The scattered light from the particle beam exits the chamber through a
standard viewport and forms an image on a translatable high-frame-rate CMOS camera (Photron SA4)
that is typically located about 350~mm outside of the chamber, see \autoref{fig:setup}.

Our illumination system consists of three different optical lasers and two different illumination
geometries. In the first configuration the full particle beam is illuminated with a collimated,
counter-propagating CW laser (Coherent Verdi V5, 532~nm, 5~W), as depicted in \autoref{fig:setup}.
The laser beam is expanded and collimated not only to illuminate the whole particle beam, but also
to avoid particle deflection~\cite{Eckerskorn:OE21:30492, Kirian:SPIE:2015,Eckerskorn:PRAppl4:064001} and damage from the
tightly focused beam. This geometry allows one to introduce a second illumination source or two
simultaneous viewing axes and an x-ray beam for diffractive imaging. The latter has been implemented
during SPI experiments at the FLASH FEL facility in Hamburg, as discussed in \autoref{sec:integration-x-ray}. In
the second illumination configuration we use a laser beam propagating perpendicular to the particle
beam direction, as show in \autoref{fig:setup}. This can be implemented alongside a counter-propagating
illumination scheme. We have utilized two short-pulse lasers, either a Nd:YLF laser (Spectra Physics
Empower ICSHG-30, 527~nm, pulse duration 100~ns, repetition rate 1~kHz, pulse energy 20~mJ,
average power 20~W), or a fiber-coupled diode laser (DILAS High-Power Diode Laser IS21.16-LC, 640~nm,
average power 10~W). The later is powered by high speed diode driver (Dr.~Heller Elektronik,
UHS-500-12.8~A, repetition rate up to 1 MHz, pulse durations $10$--$100$~ns) and mounted in oblique
orientation to maximize forward scattering. The diode laser is the least expensive option and
delivers a top-hat intensity profile.

\subsection{Transverse-plane imaging configuration}
\label{sec:transverse-imaging}
An inverted microscope is located directly below the aerosol injector to image 2D transverse beam
profiles in real time, as shown in \autoref{fig:setup}. A $5\times$ infinity-corrected objective
forms images as particles adhere to a transparent gel on a microscope slide that is manipulated with
a three-axis translation stage. A polarizing beam splitter is mounted below the microscope slide,
which allows scattered light to be imaged while the counter-propagating laser illuminates the
particle beam for side-view imaging. The entire microscope assembly is mounted on a three-axis
motorized stage so that it can be moved in and out of the interaction region during experiments, or
translated along the axis of the injector to probe the particle beam at variable distances from the
tip of the injector. In addition to producing transverse views of the particle beam, the microscope
slide is used to protect the counter-propagating laser optics (since few particles adhere to the
bare glass slide) as well as to align the laser beam to the particle beam. This alignment is done by
iteratively tilting the laser beam or translating the injector while viewing the particle/laser
overlap at two different distances (50~mm apart) along the axis of the injector.

\section{Experimental results and discussion}
\label{sec:experimental-results}
\subsection{Side-view imaging}
\begin{figure}
   \centering%
   \subfigure[]{\label{fig:time-avg-moticam}\includegraphics[width=0.45\linewidth]{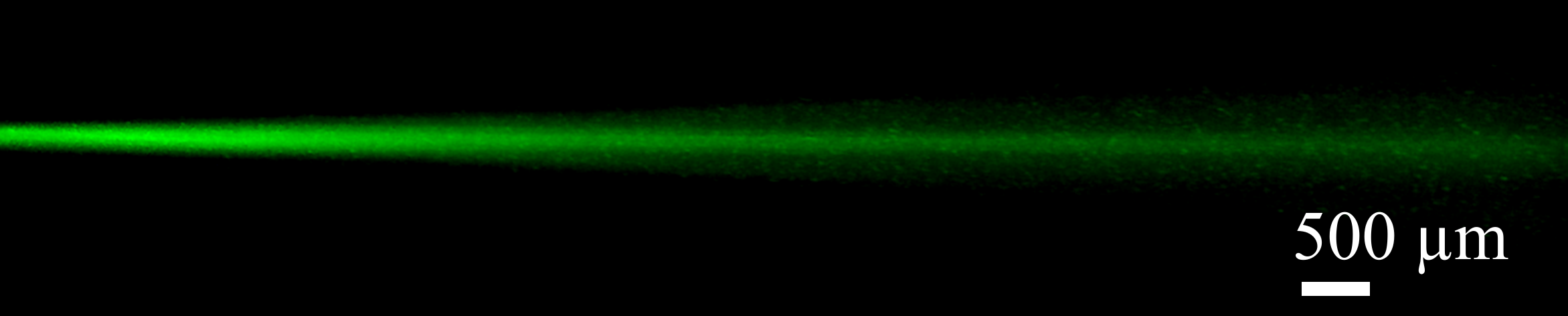}}
   \subfigure[]{\label{fig:streak-camera-shutter}\includegraphics[width=0.45\linewidth]{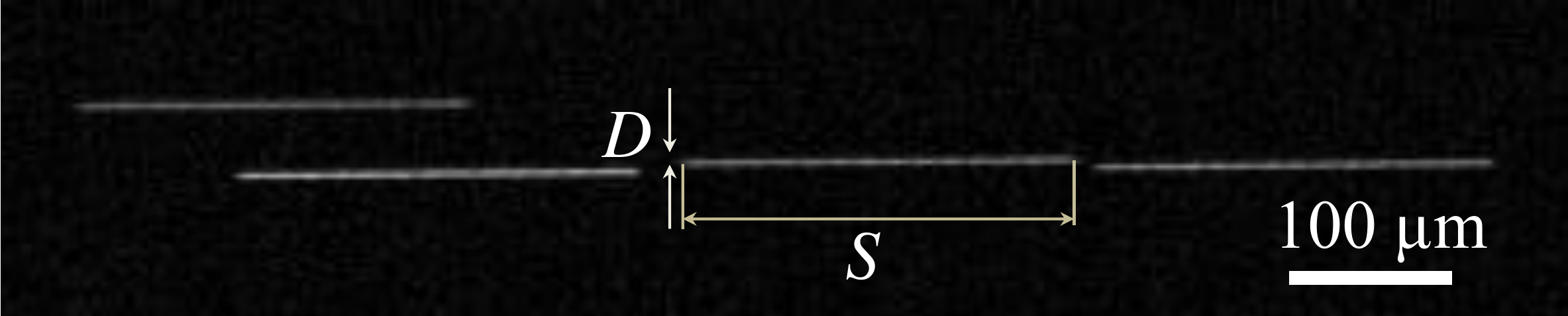}}
   \subfigure[]{\label{fig:streak-DILAS}\includegraphics[width=0.45\linewidth]{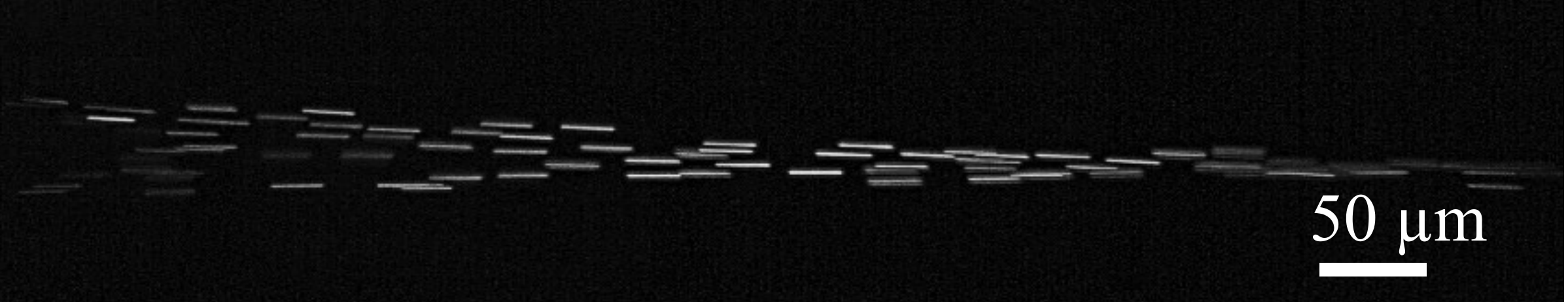}}
   \subfigure[]{\label{fig:snapshot-empower}\includegraphics[width=0.45\linewidth]{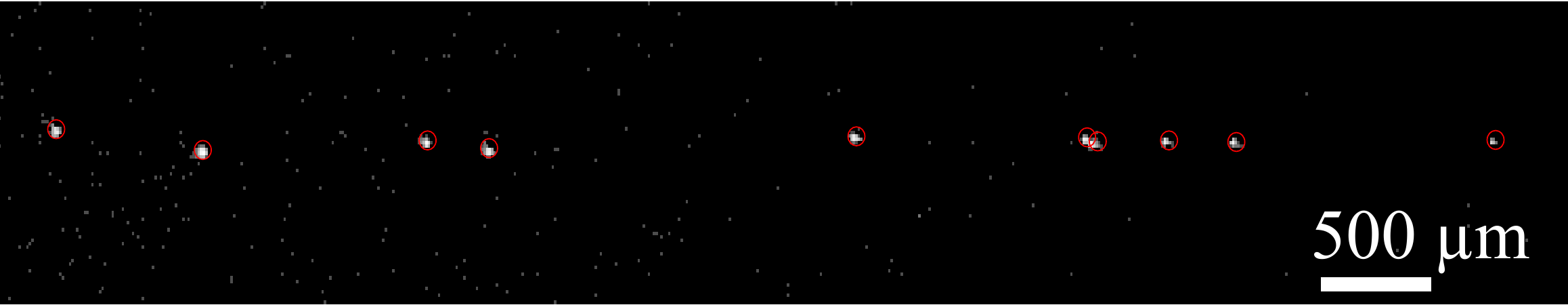}}
   \caption{Experimental particle-beam images. (a) Long-exposure image with CW laser illumination
      and 2~s camera exposure. (b) Particle-streak image of 2~\um PS moving at 18~m/s using CW laser
      illumination and a 13.5~\us camera exposure time. The length $S$ and width $D$ of the particle
      streak are marked in the image. (c) Streak image of Granulovirus (GV) particles size
      $200~\text{nm}\times200~\text{nm}\times400$~nm moving at a speed of 240~m/s, using 100~ns
      diode laser illumination and a 1~ms camera exposure time (100 pulses in the single camera
      exposure). (d) Snapshot image of 2~\um PS particles using 100~ns pulses from the Nd:YLF laser
      and 20~ms camera exposure (20 pulses in the single camera exposure). The center of the red
      circle depicts centroid of a particle snapshot.}
   \label{fig:raw-images}
\end{figure}
Representative images from our side-view imaging scheme are shown in \autoref{fig:raw-images} for
the different imaging modes introduced in \autoref{sec:imaging-schemes}.
\autoref{fig:time-avg-moticam}, \ref{fig:streak-camera-shutter}, and \ref{fig:snapshot-empower} show
images of $d=2$~\um polystyrene-sphere particles (PS), and in \autoref{fig:streak-DILAS} shows an
image of $d\approx300$~nm GV particles. The PS particles were injected with an ALS, whereas GV
particles were injected with a convergent-nozzle injector. \autoref{fig:time-avg-moticam} shows a
typical long-exposure image collected with counter-propagating CW beam illumination, which may be
interpreted as a projection of the particle beam density since the particle size distribution is
relatively narrow (< 5\%). The results of streak imaging for particles moving at two different
speeds are shown in \autoref{fig:streak-camera-shutter} and \ref{fig:streak-DILAS}.
\autoref{fig:streak-camera-shutter} shows particle streaks recorded using counter-propagating CW
beam illumination and a short integration time of 13.5~\us on the camera, while
\autoref{fig:streak-DILAS} is recorded with illumination by multiple 100-ns laser pulses and a long
camera integration time of 1~ms. A snapshot image with $t_\text{exp}<\tau$ of 2~\um PS moving at
approximately $18$~m/s is shown in \autoref{fig:snapshot-empower}. Here, short illumination times
(100~ns) and relatively slow particle speed lead to distinct single spots on the camera, highlighted
in \autoref{fig:snapshot-empower} by red circles, which are centered around the calculated centroid
positions of individual particles.

In the following we demonstrate how these data can be used to reconstruct the particle-beam density
and velocity distributions.
\begin{figure}[t]
   \centering
   \subfigure[]{\label{fig:particle-density-ALS}\includegraphics[width=0.8\linewidth]{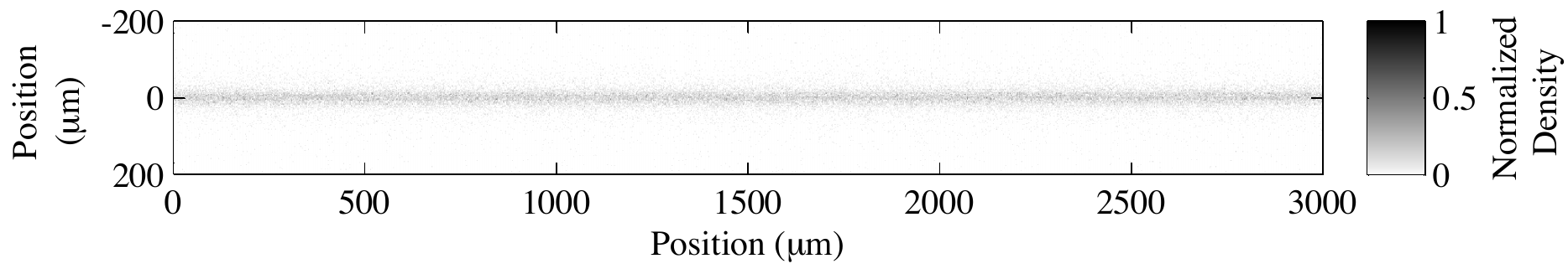}}\\
   \subfigure[]{\label{fig:particle-density-capillary}\includegraphics[width=0.8\linewidth]{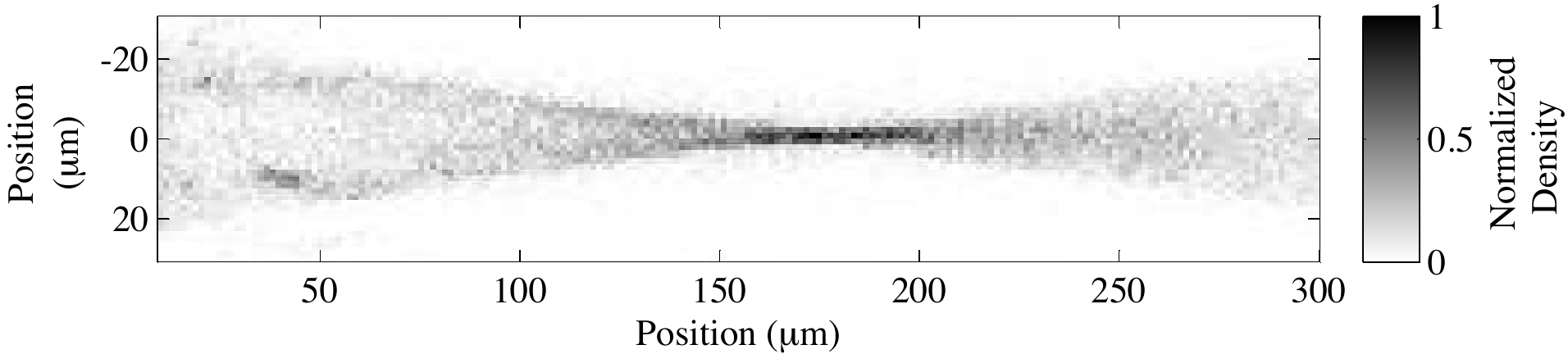}}\\
   \caption{The 2D projected particle density map of (a) 2~\um PS focused by the ALS and (b) 500~nm
      PS focused with the converging capillary injector.}
   \label{fig:particle-density}
\end{figure}
Two-dimensional particle density maps generated from raw side-view images are shown in
\autoref{fig:particle-density}. For CW illumination in the long-exposure mode, the image intensity
is directly proportional to the projected particle density, provided that only a single particle
species is present. This allows for direct monitoring of the injector behavior through the
observation of relative image intensities, but does not readily allow for a quantitative evaluation
of particle number densities without careful calibration measurements. On the other hand, streak and
snapshot imaging modes allow for direct and quantitative measurements of the particle beam density
without the need for intensity calibrations, since particle image centroids (for both streaks or
spots) can be determined with a precision better than diffraction limit
~\cite{Betzig:Science313:1642, Henriques:Biopolymers95:322}. From these centroids (seen in
\autoref{fig:snapshot-empower}), projected particle density maps can be produced, which allow
quantitative estimates of expected hit fractions in SPI experiments.

The streak imaging technique allows for the estimation of particle velocities through evaluation of
the streak length with a single pulse using a well-calibrated imaging system, as indicated in
\ref{fig:streak-camera-shutter}. A velocity measurement is also feasible using snapshot imaging if
more than one illumination pulse occurs while the particle is in the field of view, either in the
same frame or successive frames. We note several pitfalls that need to be avoided for accurate
determination of particle velocity distributions from side-view imaging measurements:
\begin{enumerate}
   \itemsep=0pt
\item The temporal illumination intensity profile of a pulsed laser source will be reflected in the
   spatial intensity of the particle streak; often one might observe long, faint trails from each
   particle, due to a slow decay of the laser pulse intensity. An accurate determination of the
   velocity requires knowledge of the temporal laser profile to disentangle the spatial image of the
   particle. Ideally, the illumination source should have a top-hat temporal profile.
\item Particles moving out of the illuminated volume or FOV during the exposure time will appear to
   produce shorter streaks. This can be avoided by ensuring the illumination to be large enough to
   cover the entire particle beam in the FOV and ignoring streaks that lead to the edge of the image
   during velocity analysis.
\item Particles that move outside of the depth-of-focus of the imaging system will result in
   de-focused images and in some cases non-uniform streaks. If not corrected for, this will result
   in systematic errors in velocity estimates. However, the inclusion of image de-focus in the
   analysis algorithm could, in principle, reveal 3D information from a single view upon careful
   calibration~\cite{Willert:ExpFluids12:353, Guerrero:MST17:2328}. For the narrow particle beams
   considered here, de-focus is typically not a significant problem and can be ignored.
\end{enumerate}
Once particle densities and velocity distributions are obtained, the injector transmission
efficiency can be determined by comparing the rate ($R_\text{in}$) at which particles enter and the
rate ($R_\text{out}$) at which they leave the injector. $R_\text{out}$ can be calculated from the
expression $n=R_\text{out}l/v$, where $n$ is the total number of particles contained within a planar
slab of thickness $l$, where particles are injected at a frequency $f$ at a velocity $v$ in the
direction normal to the slab~\cite{Kirian:SD2:041717}.

\subsection{Transverse-plane imaging}
\begin{figure}
   \centering
   \includegraphics[width=0.9\textwidth]{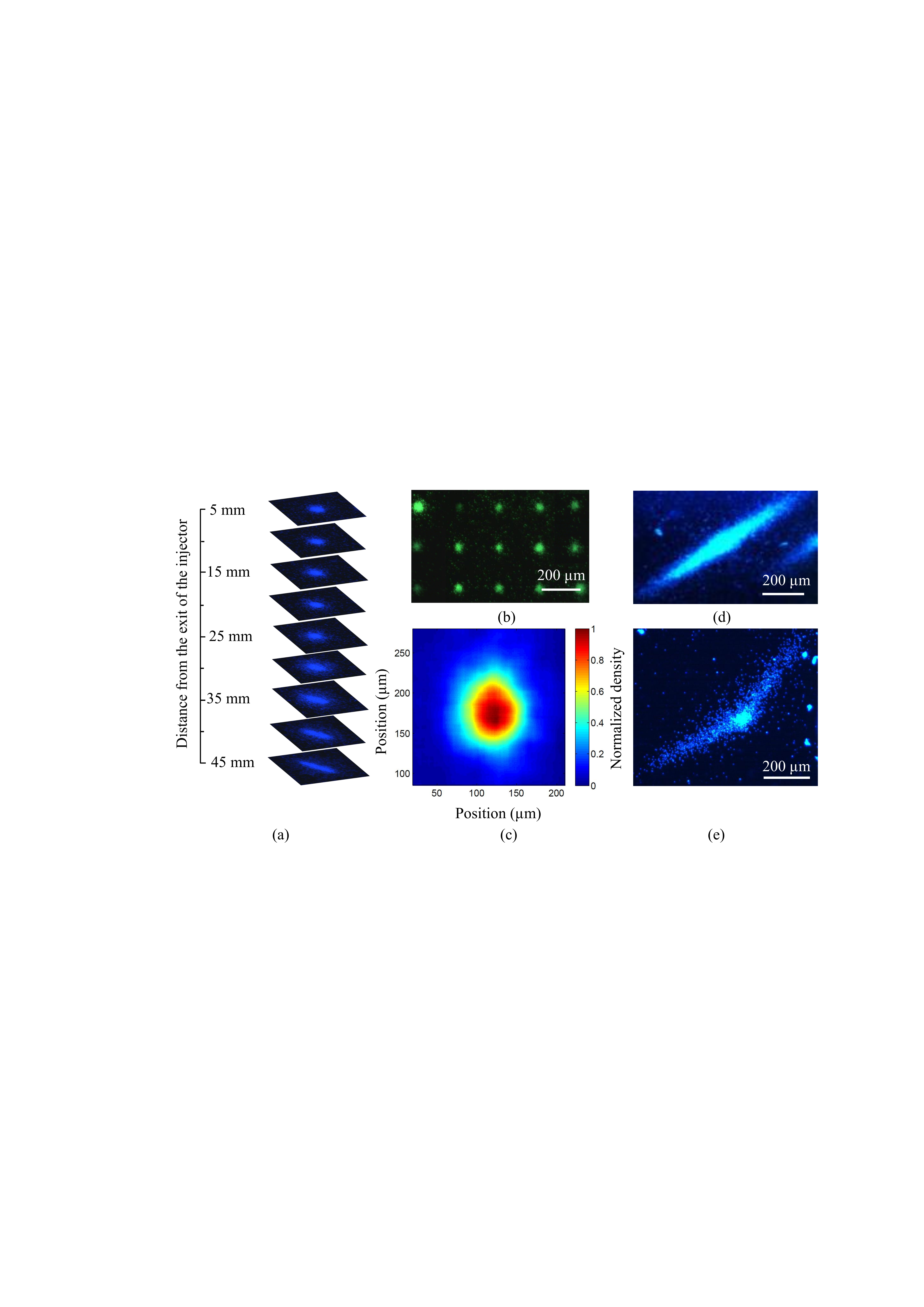}
   \caption{Imaging of 2~\um PS particles from a beam focused by an ALS injector and deposition on
      transparent gel through an in-vacuum inverted microscope. (a) A series of images recorded at
      different distances from the injector tip. (b) A lateral scan of the microscope slide at a
      distance of 35~mm from the injector tip, recording particle distributions following a 1~min
      particle deposition per spot. (c) The transverse particle-beam density profile obtained in a
      measurement similar to (b), but under conditions where individual particles could be observed
      and their centroids determined. (d, e) 1~min particle depositions from a poorly performing
      injector, which, perhaps, is caused by dispersion in particle sizes and/or asymmetry in the
      particle source at the inlet of the injector.}
   \label{fig:dustingspot}%
\end{figure}

Poorly performing injectors sometimes generate asymmetric particle beams, analogous to astigmatisms
in optical systems. This is not readily detectable in side-view imaging configurations, but is
clearly visible through transverse-plane imaging with the inverted microscope discussed previously.
\autoref{fig:dustingspot} shows particle-deposition images at different distances from the tip of an
ALS injector, and a clear variation in particle beam asymmetry with position. For particles larger
than 1~\um, individual particles can be detected as they adhere to the gel, allowing
semi-quantitative analysis of the particle beam width on the transverse plane in real time, as shown
in \autoref{fig:dustingspot}\,(c).
However, the accuracy of the analysis is limited by our understanding of how particles adhere to the
gel surface -- most importantly, how the likelihood of particle adherence changes with time, \eg, as
particles accumulate. For particles on the order of 100~nm or smaller, a detectable
particle-deposition image is obtained after a few seconds of accumulation time, depending on the
concentration and size of particles.

\subsection{Injector optimization}
\begin{figure}
   \centering
   \subfigure[]{\label{fig:VelocityParticleDensityvsPressure}\includegraphics[width=0.54
   	\linewidth]{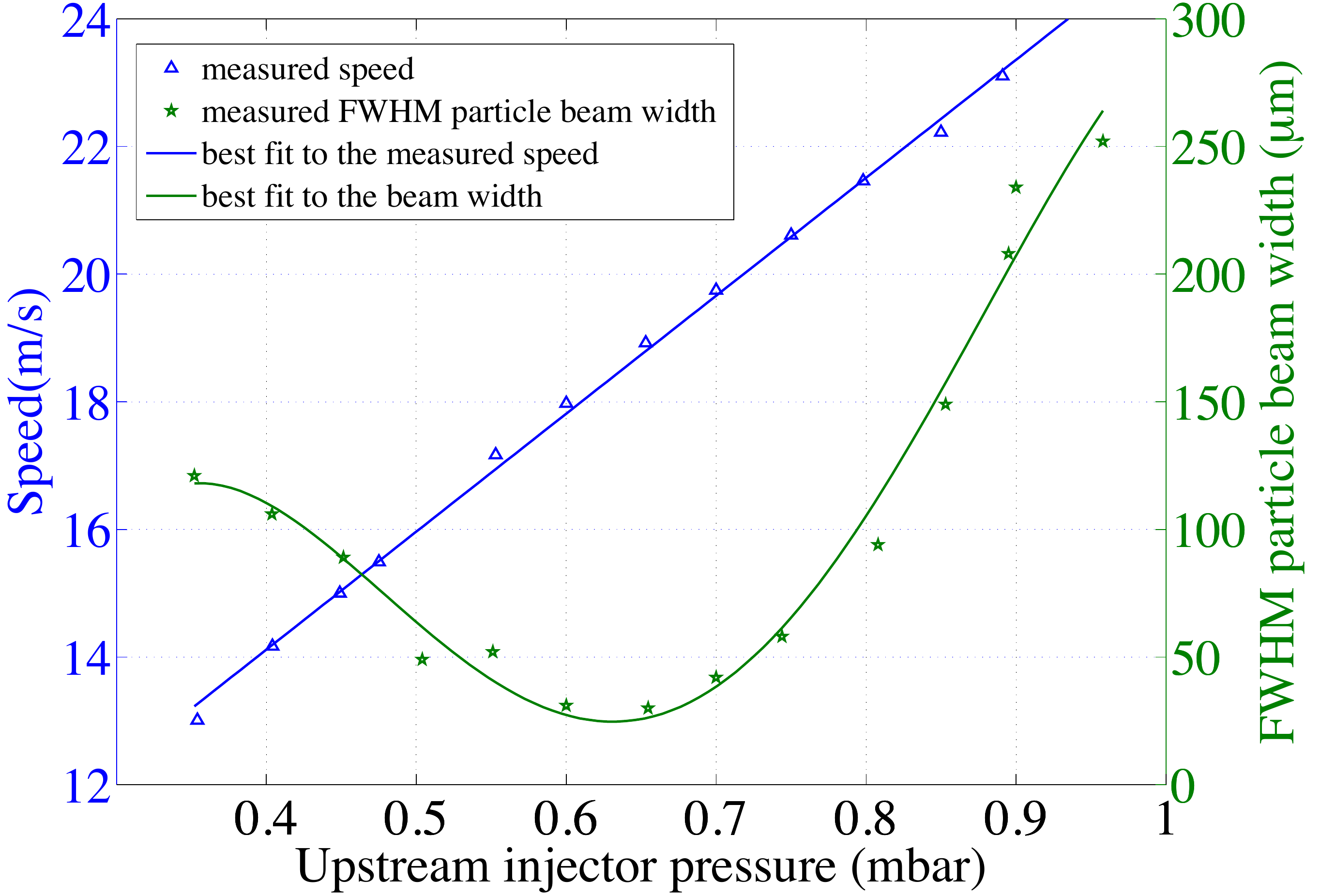}}
   \subfigure[]{\label{fig:VelocityHist}\includegraphics[width=0.39\linewidth]{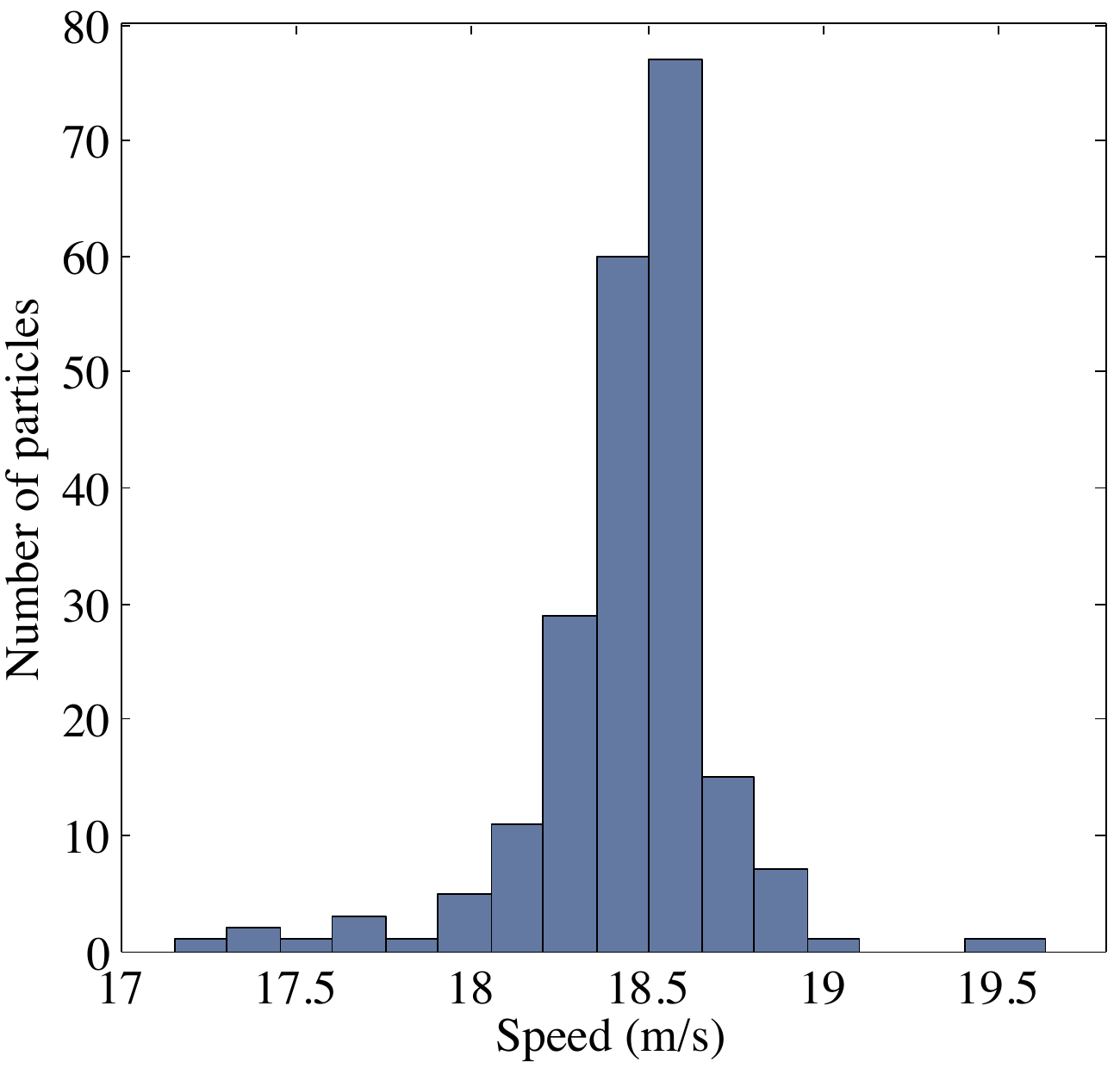}}
   \caption{Injector-performance measurement for 2~\um PS particles focused by the aerodynamic-lens-stack
      aerosol injector. (a) Particle velocities and projected density \emph{versus} injector
      upstream pressure. (b) Velocity distribution of the particles at 0.65~mbar upstream pressure.}
   \label{fig:hit-rate}
\end{figure}

The presented characterization methods offer a powerful means to optimize the performance of
particle injectors, both \emph{online} during SPI measurements at XFEL facilities, as well as
offline in the preparation laboratory. As discussed in \autoref{sec:TheoryBackground}, the 2D
projected number density of the particle beam is the most important parameter that needs to be
optimized, since it scales directly with the hit fraction in an SPI experiment. The hit fraction
depends on particle velocity, injector transmission, and particle beam diameter, which are ideally
measured independently while developing and optimizing aerosol injectors.
\autoref{fig:VelocityParticleDensityvsPressure} shows a typical plot for injector optimization,
including the velocity and particle-beam diameter in the case of 2-\um PS particles measured 35~mm
downstream from the tip of the injector, as a function of the upstream pressure of the injector. The
downstream chamber pressure is maintained below $10^{-2}$~mbar and does not significantly effect the
particle speed or beam diameter. The velocity increases linearly with the upstream pressure.
However, the particle beam size exhibits a distinctive minimum around 30~\um FWHM, at an upstream
pressure of 0.63~mbar. As seen from \autoref{fig:VelocityHist}, the particles are moving at an
average velocity of 18.49~m/s with standard deviation of 0.28~m/s at this upstream pressure.
Ignoring the transmission efficiency for now, the optimum operating pressure of the injector for
maximum hit fraction should be chosen such that the product of these two parameters is minimized
(see \autoref{sec:TheoryBackground}), \ie, for 2~\um PS particles, the injector should be operated
at 0.6~mbar for maximum hit fraction. Alternatively, one may simply measure the projected particle
beam density, which automatically accounts for the contributions of velocity, transmission
efficiency, and particle beam diameter. In practice, the assumption of a constant transmission
efficiency is not valid over a large pressure range, and this needs to be taken into account. We
note that using streak or multiple-exposure snapshot imaging allows a quantitative measure of the
transmission efficiency.

\subsection{Integration with x-ray experiments}
\label{sec:integration-x-ray}
\begin{figure}
   \centering
   \includegraphics[width=0.6\textwidth]{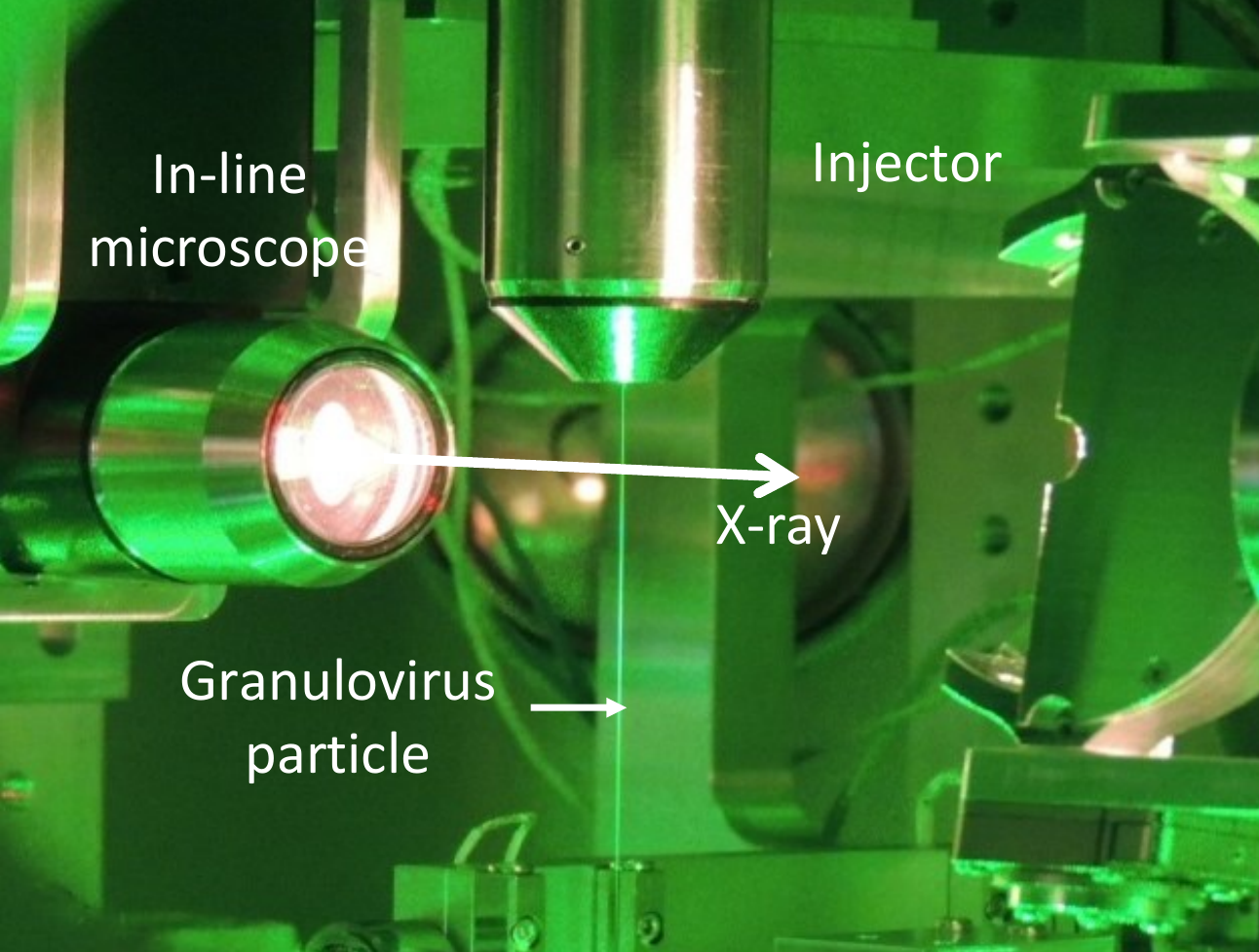}
   \caption{The SPI setup at FLASH. The photograph of the setup shows the space around the
      interaction region of the laser beam with a stream of GV. This photo was taken by a DSLR
      camera through a window on the experimental vacuum chamber.}
   \label{fig:FLASH-setup-photo}
\end{figure}

We demonstrated the utility of optical particle-beam imaging using a custom SPI experimental
apparatus at the FLASH free-electron laser facility in Hamburg, Germany. Due to space limitations,
we utilized a counter-propagating 5W CW laser, as shown in \autoref{fig:FLASH-setup-photo}. An
in-line microscope with a long-exposure CCD was placed on the same axis as the x-rays, in addition
to a high-speed camera that imaged the particle beam from a viewpoint perpendicular to the x-ray
beam axis. This enabled us to have, simultaneously, two orthogonal side views of the particle beam.
The long exposure images from the in-line microscope were used to position of the injector for
maximum hit rate, whereas images from the high-speed camera were used to position the beam with
respect to the x-ray focus and to provide real-time estimates of the particle velocity and number
density. As seen from \autoref{sec:SideImaging}, the counter-propagating illumination scheme leaves
plenty of spaces around the interaction region for multiple views and additional diagnostics.
However, it requires careful alignment of the laser with the particle beam, especially for the case
of CW lasers that must be focused to smaller diameters (approximately 100~$\mu$m in this particular
case) than pulsed lasers of equivalent average power. We therefore installed translation and tilting
stages inside the vacuum chamber for steering and translating the laser beam. In order to mask the
scattering light from the injector tip we constructed a light shielding around the objective lens.
As seen in \autoref{fig:FLASH-setup-photo}, once the CW laser is properly aligned, the average
intensity from a beam of GV particles is easily visible to a typical CCD (in this case, a consumer
single-lens reflex camera). The ability to immediately see a particle beam drastically reduced the
time needed to align the injector, and immediately revealed the typical fluctuations in the injector
transmission efficiency.

\section{Summary and conclusion}
We demonstrated the utility of direct optical imaging of micro- and nano-particle aerosol beams for
the purpose of improving the overall efficiency of single-particle x-ray diffractive imaging (SPI)
experiments. We find that direct imaging of the particle beam is a straightforward means to
quantitatively measure particle density maps, particle velocity distribution, and injector
transmission efficiency, which are key diagnostics for optimizing SPI experiments at large-scale
x-ray facilities where the time available for measurements is rather limited. A modest setup with an
off-the-shelf CW laser of $\sim$1--5~W power can readily reveal the time-averaged position and width
of a typical particle beam, which greatly simplifies the procedure of positioning the injector with
respect to the x-ray beam. The overall brightness of the particle beam is also indicative the
injector performance. Remarkably, such a simple diagnostic can save many hours of effort, and
corresponding facility costs, compared to ``shooting blind'', \ie, when injection is optimized based
on x-ray diffraction data. We also showed that pulses of
well-defined duration as well as a CW laser combined with a camera with a fast shutter can
simultaneously produce quantitative particle-density and velocity-distribution maps. Our side-view
imaging schemes were complemented by a compact in-vacuum microscope that enables indirect particle
beam imaging in the transverse plane, which readily reveals particle-beam astigmatism that is not
easily observed from viewpoints that are orthogonal to the particle beam.

In the configurations considered here, we have also imaged individual $200$~nm diameter particles
moving at speeds of $300$~m/s~\cite{Kirian:SD2:041717} with a modest short-pulse laser ($100$~ns and
10~W average power). Simple scattering estimates suggest that much smaller particles, perhaps down
to few tens of nanometers, should also be visible with a sufficiently intense illumination
(approximately 100~mJ pulses focused to about 1~mm diameter) and a very-high-sensitivity imaging
device. Although velocity measurements can be made from short pulses that create streaked particle
images, it appears that the optimal method for determining velocities, from a signal-to-noise
standpoint, is through the use of two time-delayed pulses of duration short enough to produce
``snapshot'' diffraction-limited particle images. Pulse durations of approximately 5~ns are required
to freeze the motion of particles moving at 200~m/s for an image resolution of 1~$\mu$m, but
Q-switched lasers that produce such pulses are common and relatively inexpensive.

We tested three different imaging modes that differ in terms of illumination geometry, optics, and
the illumination source. Each of them can be implemented relatively straightforwardly in typical SPI
experiments with only minor modifications. A counter-propagating geometry, in which the particle and
laser beams oppose each other, maximizes the space available for ancillary diagnostics such as
time-of-flight spectrometers, but requires a transparent shield to maintain clean beam-steering
optics below the injector and unnecessarily exposes upstream particles to laser illumination. For
imaging the smallest of particles, it may become necessary to operate above the damage threshold of
the particles. Hence, a transverse illumination scheme would be required to avoid damaging particles
prior to probing with x-rays.

Thus far, our apparatus has been used to characterize the injection process downstream of the
injector, close to the interaction region of particles and x-rays. It would be advantageous to
include similar imaging diagnostics at positions upstream of the injector exit, so that the aerosol
formation and pre-collimation (prior to focussing) can also be monitored and de-coupled from the
downstream particle-beam focusing components. Ideally, these diagnostics would be extended to
include particle size measurements through careful calibrations of integrated scattering intensity,
Mie scattering profiles, or other interferometric methods. Such \emph{in-situ} measurements would
allow us to monitor particle aggregation and evaporation rate of the liquid buffer from the initial
droplets generated by the nebulization device~\cite{Vervoort:JCA1189:92}.

\section*{Acknowledgments}
In addition to DESY, this work has been supported by the excellence cluster ``The Hamburg Center for
Ultrafast Imaging—Structure, Dynamics and Control of Matter at the Atomic Scale'' of the Deutsche
Forschungsgemeinschaft (CUI, DFG-EXC1074) and by the Australian Research Council's Discovery
Projects funding scheme (DP110100975). R.A.K.\ acknowledges support from an NSF STC Award (1231306).
J.K.\ acknowledges support by the European Research Council through the Consolidator Grant COMOTION
(ERC-614507).
\end{document}